\documentclass{article}

\usepackage[preprint, nonatbib]{neurips_2026}
\usepackage[numbers]{natbib}

\usepackage{amsmath, amssymb, amsfonts}
\usepackage{graphicx}
\usepackage{booktabs}
\usepackage{nicefrac}
\usepackage{microtype}
\usepackage{xcolor}
\usepackage{enumitem}
\usepackage{array}
\usepackage{multirow}
\usepackage{subcaption}
\usepackage{algorithm}
\usepackage{algorithmic}
\usepackage{rotating}

\usepackage[utf8]{inputenc}
\usepackage[T1]{fontenc}
\usepackage{hyperref}
\usepackage{url}

\DeclareMathOperator{\pt}{\mathnormal p_\mathrm{T}}
\DeclareMathOperator{\MET}{\mathnormal E_\mathrm{T}^\mathrm{miss}}
\DeclareMathOperator{\logp}{\mathnormal{\log p(x^\prime, m)}}

\newcommand{\LQ}{$LQ \rightarrow b\tau$}
\newcommand{\Allll}{$A \rightarrow 4l$}
\newcommand{\hzero}{$h^0 \rightarrow \tau\tau$}
\newcommand{\hplus}{$h^+ \rightarrow \tau\nu$}

\title{Distilling Normalizing Flows for Real-Time Anomaly Detection at the LHC}

\author{
Tara P. A. Tahseen$^{1,3}$\thanks{These authors contributed equally to this work.} \quad Noah Clarke Hall$^{2*}$ 
\\
\textbf{Nikolaos Konstantinidis$^1$ \quad Paula Martínez Suárez$^{2,3}$}\\
$^1$Department of Physics and Astronomy, University College London, WC1E 6BT, UK \\ $^2$Experimental Physics, CERN, Espl. des Particules 1, Meyrin, 1217, Switzerland \\
$^3$CERN openlab, CERN, Espl. des Particules 1, Meyrin, 1217, Switzerland\\
\texttt{tara.tahseen.22@ucl.ac.uk}\\
\texttt{noah.clarke.hall@cern.ch}\\
}

\begin{document}

\maketitle

\begin{abstract}
Normalizing flows are principled anomaly detectors, selecting anomalies using a probabilistic per-event likelihood. 
Extreme latency and resource constraints have prevented the deployment of flow likelihoods within the hardware triggers at the Large Hadron Collider. 
We bypass these limitations by distilling the likelihood from a large normalizing flow into lightweight student estimators suitable for deployment on a field-programmable gate array. 
Our use of a conditional normalizing flow enables precise likelihood estimation in the presence of missing input features.
Both decision tree and neural network students are considered, the latter optimized under advanced quantization techniques. 
By simultaneously improving likelihood quality and lowering inference cost, we demonstrate both state-of-the-art physics performance and latency compared with existing flow-based approaches.

\end{abstract}

\section{Introduction}

The Large Hadron Collider (LHC) produces collisions at 40~MHz, far more than can be stored. A hardware Level-1 Trigger (L1T) performs the first stage of data reduction, implemented on field-programmable gate arrays (FPGAs) with a latency budget of a few microseconds. 
Unsupervised anomaly detection offers an unbiased approach to event selection, selecting outliers in the absence of labelled examples. Deploying such a model at the L1T requires inference in $\mathcal{O}(100~\text{ns})$ using a small fraction of the FPGA resources~\cite{Govorkova_2022}. 

The ATLAS~\cite{ATLAS} and CMS~\cite{CMS} experiments have deployed several unsupervised anomaly detection algorithms at the L1T level~\cite{Gandrakota:2025X5, CMS-DP-2024-121, pol2023knowledge, Sugizaki:2941530, MartinezSuarez:2963506}. 
All are based on autoencoder neural networks, scoring events using an input-output reconstruction error or latent coordinates.
Autoencoders based on decision trees~\cite{Roche2024} and spiking networks~\cite{Dillon_2026} have also been proposed for L1T deployment.
Normalizing flows provide an attractive alternative, scoring events using the exact likelihood $p(x)$ \citep{papamakarios2021normalizing} of the event occurring within the training set. 
However, exact computation of flow likelihoods on FPGAs is impractical, requiring the numerical integration of a neural vector field (continuous flows, \cite{Vaselli_2026}) or the evaluation of complex sequential transformations and associated Jacobians (discrete flows, here). 
A prior study by \cite{Vaselli_2026} addressed this by replacing the exact likelihood with a cheaper proxy such as the vector field itself. 
While L1T-compatible thanks to advanced quantization techniques, this approach was found to be less anomaly-efficient than the full likelihood.

In the work presented here, we seek to bypass the efficiency-cost tradeoff using knowledge distillation, regressing the outputs of a large, FPGA-incompatible ``teacher" using a lightweight, FPGA-compatible ``student".
This approach was originally proposed for anomaly detection at the LHC by~\citet{pol2023knowledge}, and implemented for autoencoders by \cite{CMS-DP-2024-121, Gupta:2942542}.
In the absence of FPGA-compatibility constraints, the teacher can be large and heterogeneous and leverage feature engineering in order to model the likelihood with maximal precision.
Our use of a conditional normalizing flow enables precise likelihood estimation in the presence of missing input features, a challenge not addressed by \cite{Vaselli_2026}.
Likelihood precision is closely aligned with anomaly detection, since incorrectly assigning high likelihoods to normal events will increase the rate of false positives, and vice versa.
We regress the flow likelihood directly from detector inputs using both a boosted decision tree (BDT) and a dense neural network (DNN).
The resulting FPGA implementations improve anomaly discrimination and lower latency compared with previous flow-based approaches.

\section{Datasets}
We use the datasets employed by \citet{Govorkova_2022}, published on Zenodo \citep{thea_aarrestad_2021_5046389_sm, thea_aarrestad_2021_5055454_lq, thea_aarrestad_2021_5046446_a4l, thea_aarrestad_2021_5061688_hc, thea_aarrestad_2021_5061633_h0}, and discussed in detail in \cite{lhc_dataset}. 
The dataset contains simulated and pre-filtered proton–proton collisions containing an electron or a muon with a transverse momentum $\pt > 23$ GeV and pseudorapidity $|\eta| < 3$ (electron) and $|\eta| < 2.1$ (muon). 
This selection is representative of the L1T of a multipurpose LHC experiment. 
The Standard Model (SM) data sample represents a typical proton–proton collision dataset \citep{thea_aarrestad_2021_5046389_sm} and contains 13.5 million events. 
The SM dataset is split 40-10-50 into training, validation, and testing datasets.
An additional four new-physics scenarios are benchmarked:

\begin{itemize}
    \item  A leptoquark with a mass of 80~GeV, decaying to a $b$ quark and a $\tau$ lepton (\LQ) \citep{thea_aarrestad_2021_5055454_lq},
\item A neutral scalar boson with a mass of 50~GeV, decaying to two off-shell $Z$ bosons, each forced to decay to two leptons (\Allll) \citep{thea_aarrestad_2021_5046446_a4l},
\item A scalar boson with a mass of 60~GeV, decaying to two $\tau$ leptons (\hzero) \citep{thea_aarrestad_2021_5061633_h0},
\item  A charged scalar boson with a mass of 60~GeV, decaying to a $\tau$ lepton and a neutrino (\hplus) \citep{thea_aarrestad_2021_5061688_hc}.
\end{itemize}

Each event is represented by the $(\pt, \eta, \phi)$ for 18 reconstructed particles (4 electrons, 4 muons, and 10 jets) alongside the magnitude and $\phi$-coordinate of the missing transverse energy ($\MET$).
The $\MET$ $\phi$ coordinate is subtracted from each $\phi$, and the now-redundant $\MET$ $\phi$ is removed.
The momenta of missing particles are represented by imputing zeros.
The resulting vector, $x$, of unnormalized particle-level features has $55$ components.

\section{Normalizing flow}
Feature engineering is applied on $x$ in order to make the density of SM events easier to learn.
The decomposition $\phi\rightarrow(\sin\phi,\cos\phi)$ is taken in order to capture $\phi$ periodicity.
The transverse momenta transform under $\pt\rightarrow \log (1+\pt)$ to suppress the variance introduced by high-$\pt$ tails.
The integer multiplicities of muons, $n_\mu$, electrons, $n_e$ and jets, $n_j$, are appended to the input vector.
The engineered features consist of 73 normalized particle-level features ($x^\prime$) plus 3 multiplicities ($m$).
All features are scaled to unit variance using scales measured on the SM training dataset, not including imputed zeros.
Imputed zeros in $x^\prime$ are replaced by a small dequantization noise $\epsilon\sim\mathcal{N}(0, 10^{-2})$.

A key challenge for improving the learned likelihood is the treatment of missing data.
Normalizing flows are not naturally suited to learning sparse representations, since they cannot directly model data on a lower-dimensional manifold of unknown dimension \citep{NEURIPS2021_dfd78699}.
As a result, using a normalizing flow to model $p(x)$ or $p(x^\prime, m)$ directly is unlikely to produce precise likelihood estimates, especially if model capacity is limited by FPGA compatibility as in \cite{Vaselli_2026}.
We resolve this by using a conditional normalizing flow \citep{kingma2019introduction} to model the conditional likelihood $p(x^\prime|m)$.
Conditioning on $m$ guarantees the tractability of the modeling task by breaking down the target distribution into a mixture of low-dimensional continuous distributions.
Since every possible combination of $(n_\mu, n_e, n_j)$ can be assigned an integer index, the prior $p(m)$ can be modeled using a categorical distribution.
Combining the conditional flow and prior, we recover the joint log-likelihood
\begin{equation}
    -\logp = -\underbrace{\log p(x^\prime|m)}_\text{flow} - \underbrace{\log p(m)}_\text{prior}
\end{equation}

The flow uses the rational-quadratic spline bijectors proposed by \citet{durkan2019neuralsplineflows} and implemented in the \texttt{nflows} package~\cite{nflows}.
The flow consists of 10 spline transformations. Each spline has 3 knots, with the knot locations and gradients controlled by an $m$-conditioned ResNet \citep{he2015deepresiduallearningimage} with $2$ residual blocks and $128$ nodes per hidden layer. 
Each hidden layer uses the ReLU activation, with batch normalization and a dropout probability of $0.1$.
The flow is trained for 500 epochs with a batch size of 4096.
The Adam optimizer is used with a learning rate of $10^{-4}$.
In total, the flow has 1.3~M trainable parameters.
Classifying between flow samples and SM data using a BDT, we measure the cross-entropy to be 0.68, indicating that the learned density is almost indistinguishable from the SM distribution.

\section{Knowledge distillation}
Since $-\logp$ is the result of both a normalizing flow and a prior, it is more challenging to define an FPGA-compatible score such as the vector field $v_t$ used in \citet{Vaselli_2026}.
Instead, we seek to transfer $-\logp$ into an FPGA-compatible student $s(x)$ using knowledge distillation.
Regressing $-\logp$ directly from $x$ incorporates the feature engineering into the student, avoiding the need to preprocess features on the FPGA.
The students are trained by minimizing the weighted square error
\begin{equation}
    \mathcal{L}_\text{student} = w\times||s(x) + \logp ||^2,
\end{equation}
where $w$ is a normalized importance weight such that $w \propto 1/\sqrt{p(x^\prime, m)}$.
Importance weighting improves the precision of the student at extreme anomaly scores, removing the need to supervise the student on signals, in contrast to ``outlier exposure" distillation methods~\cite{CMS-DP-2024-121, pol2023knowledge}.

Two student model architectures are evaluated: a boosted decision tree (BDT) and a dense neural network (DNN).
The BDT is implemented using \texttt{xgboost}~\cite{xgboost} with a learning rate of $0.3$, a maximum ensemble size of $100$ and a maximum tree depth of $3$.
Post-training quantization (PTQ) is performed using \texttt{conifer} \citep{conifer} to compress the BDT in preparation for implementation on an FPGA.
The minimum precision needed was found to be 18 bits for the inputs and tree nodes, and 14 bits for the leaves.
The DNN 
consists of 2 hidden layers of 32 nodes with a ReLU activation.
Input normalization is incorporated into the DNN by applying batch normalization before the first hidden layer.
Quantization-aware training is performed using the HGQ method \citep{Sun:2951900} to compress the DNN in preparation for FPGA implementation.
The DNN is trained for 500 epochs with a batch size of 4096.
The Adam optimizer is used with a learning rate of $10^{-4}$ and the gradient norm clipped to 1.
The targets for the DNN are scaled to unit variance, which stabilizes training while preserving ranking in $-\logp$.
The BDT produces an (un)weighted $R^2$ score on the validation set of 0.86 (0.90), compared with 0.84 (0.95) for the DNN.




\section{Physics performance}
\begin{figure*}
    \centering
    \includegraphics[width=\textwidth]{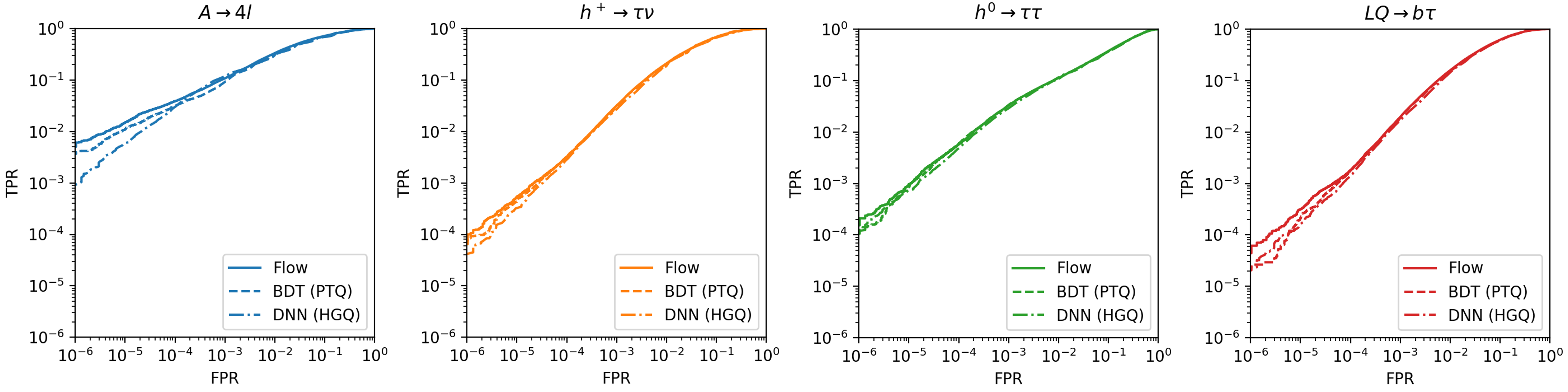}
    \caption{ROC curves for the four new-physics benchmarks, comparing the flow teacher score $-\logp$ with its BDT and DNN student surrogates.}
    \label{fig:roc_curves}
\end{figure*}

Model physics performance is evaluated using the four new-physics scenarios.
We quantify anomaly detection performance by plotting the receiver operating characteristic (ROC) curve for each model, shown in Figure~\ref{fig:roc_curves}.
We quote the area under the curve (AUC) as well as the true-positive rate (TPR) at a fixed false-positive rate (FPR) of $10^{-5}$.
The AUC measures the overall discrimination power of a given model, whereas the TPR measures discrimination power in the extreme tails of the SM score distribution.
The AUC and TPR are compared to relevant results obtained by \cite{Vaselli_2026} in Table~\ref{tab:comparison}.

The $-\logp$ from our conditional normalizing flow achieves the highest AUC for all new physics benchmarks.
Both the BDT and DNN produce AUCs within $\leq1\%$  of the teacher model.
Crucially, the teacher and both students produce AUCs that exceed those obtained by \cite{Vaselli_2026}.
Our flow has the highest TPR for \LQ, \hplus\ and \hzero\ but the lowest TPR for \Allll.
The TPRs from the BDT and DNN fluctuate about that of the teacher model, with the BDT producing consistently higher TPRs than the DNN.
We note that measuring the TPR in the extreme tails of the score distribution makes the TPR highly sensitive to effects from finite statistics, feature engineering, random initialization, and hyperparameter choice.

\begin{table*}[ht]
\centering
\setlength{\tabcolsep}{4pt}
\resizebox{\textwidth}{!}{
\begin{tabular}{llcccccccc}
\toprule
\multirow{2}{*}{Model} & \multirow{2}{*}{AD score} 
& \multicolumn{4}{c}{TPR @ FPR $10^{-5}$ [\%]} 
& \multicolumn{4}{c}{AUC [\%]} \\
\cmidrule(lr){3-6} \cmidrule(lr){7-10} & & \LQ & \Allll & \hplus & \hzero & \LQ & \Allll & \hplus & \hzero \\
\midrule

\multirow{3}{*}{Flow from \cite{Vaselli_2026}}  & ODE
& 0.04 & 3.8 & 0.04 & 0.05 
& 80 & 88 & 86 & 69 \\

& $v_t$
& 0.04 & 2.8 & 0.04 & 0.06 
& 80 & 82 & 84 & 68 \\

& $v_t$ (HGQ)
& 0.04 & 3.4 & 0.05 & 0.06 
& 77 & 86 & 82 & 66 \\

\midrule
\multirow{3}{*}{Flow (ours)} &  $-\logp$ & 0.05 & 1.5& 0.05& 0.10& 88& 90& 90& 76\\
& BDT (PTQ) & 0.04& 1.1& 0.05 & 0.10& 88 & 89& 89 & 76\\
& DNN (HGQ) & 0.04 & 0.58 & 0.03 & 0.07 & 88& 89& 89& 76\\
\bottomrule
\end{tabular}
}
\caption{Comparison of AUC and TPR for each model, with comparison to prior work in \cite{Vaselli_2026}.}
\label{tab:comparison}
\end{table*}

\section{FPGA implementation}
The BDT and DNN are converted to firmware using the \texttt{conifer} and \texttt{da4ml} bridges respectively.
We target a Xilinx Virtex UltraScale+ FPGA with a 200~MHz clock, using the same part number as \citet{Vaselli_2026} to ensure consistency.
Resource and timing estimates, displayed in Table \ref{tab:resources}, are obtained via RTL synthesis in Vitis HLS.
The DNN student performs similarly to the HGQ implementation of $v_t$, demonstrating that the large size of the teacher does not compromise the FPGA compatibility of the student.
The BDT student can be evaluated in just a single clock cycle thanks to the unrolling of the decision tree structure \citep{conifer}.
The latency of both the BDT and DNN students is less than that demonstrated previously by both autoencoder \cite{Govorkova_2022} and flow-based \cite{Vaselli_2026} approaches.
All the resource usages quoted represent $<1\%$ of the resources of the whole FPGA.
\begin{table*}[h!]
    \centering
    \begin{tabular}{lccccc}
    \toprule
    Model & DSP & LUT & FF & Latency [ns] & II [clk] \\
    \midrule
    $v_t$ (HGQ) from~\citep{Vaselli_2026}     &  28 & 5,798 & 1,683 & 35 & 1\\
    \midrule
    BDT (PTQ) &  0  & $12,869$ & 260 & 5 & 1\\
    DNN (HGQ) &  0  & 6,633 & 1,798 & 25  &  1 \\
    \bottomrule
    \end{tabular}
    \caption{Resource and timing estimates for a Xilinx Virtex UltraScale+ FPGA with a 200~MHz clock.}
    \label{tab:resources}
\end{table*}

\section{Conclusion}

We demonstrate that knowledge distillation enables large conditional normalizing flows to serve as high-performance anomaly detectors while remaining deployable under strict hardware constraints. By training lightweight surrogate models on the flow likelihood, we transfer its discrimination power into FPGA-compatible boosted decision tree and neural network students. The conditional flow improves likelihood modeling in the presence of missing features and achieves strong physics performance, reaching AUCs of up to 90\% on benchmark signals. The students reproduce the teacher AUC to within 1\% while achieving nanosecond-scale inference latency (5–25 ns) on FPGA hardware. This approach delivers state-of-the-art likelihood-based anomaly detection with a practical path to real-time deployment within LHC hardware triggers.

\newpage

\section*{Acknowledgements}
We gratefully acknowledge the support of the UK's Science and Technology Facilities Council (STFC). TT is supported by the STFC UCL Centre for Doctoral Training in Data Intensive Science (ST/W00674X/1) and funding from CERN openlab during a 6-month work placement at CERN. The support and collaboration of CERN openlab are gratefully acknowledged. TT thanks R. P. Nathan for helpful and interesting discussions around anomaly detection. 
This work has been partially funded by the Eric \& Wendy Schmidt Fund for Strategic Innovation through the CERN Next Generation Triggers project under grant agreement number SIF-2023-004.


\bibliographystyle{unsrtnat}
\bibliography{references}

\end{document}